\documentclass[twocolumn,prl,aps]{revtex4}
\usepackage{graphicx}

\begin{document}

\title{Decision at the beam-splitter, or decision at detection, that is the question}

\author{Antoine Suarez}
\address{Center for Quantum Philosophy\\Berninastrasse 85, 8057 Zurich/Switzerland\\suarez@leman.ch, www.quantumphil.org}

\date{April 15, 2013}

\begin{abstract}

I argue that nonlocal decision of the outcomes at detection excludes \emph{any alternative theory} to quantum mechanics: Not all that matters for the results of physical experiments is content in space-time, but all that is in space-time is accessible to observation.

\end{abstract}


\maketitle

An experiment demonstrating single-photon nonlocal decision of outcome at detection has been presented and discussed in \cite{gszgs, as12}. This experiment highlights that the \emph{standard quantum mechanics} (also called the Copenhagen or orthodox interpretation) is actually characterized by the following two principles:

\emph{Principle A:} All that is in space-time is accessible to observation, unless in case of space-like separation.

\emph{Principle Q}: Not all that matters for the physical phenomena is content in space-time.

Alternative theories assume decision at the beam splitter and deny these two \emph{Principles A and Q}.

In the follow I show that nonlocal decision at detection allow us to falsify alternative theories in a natural and uncomplicated way.
\vspace{0.2cm}

\noindent \textbf{The ``collapse of the wave function''}.\textemdash A quantum experiment always consists of a device (beam-splitter, polarizer, Stern-Gerlach, etc.) exhibiting at least two output ports, each of them monitored by a corresponding detector. With this experimental configuration, after leaving the source a ``classical'' particle can reach each detector by two different paths.

\begin{figure}[t]
\includegraphics[width=80mm]{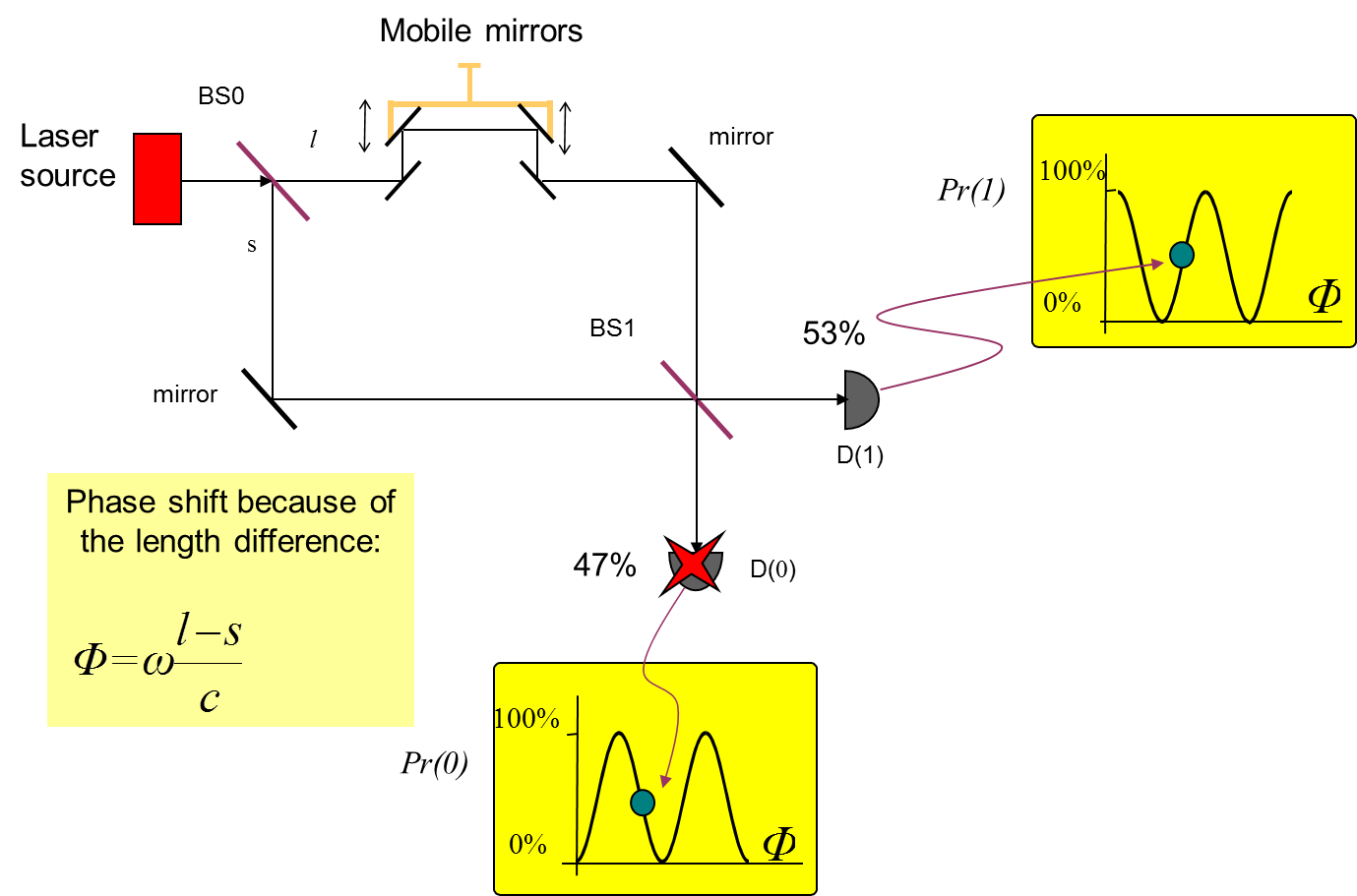}
\caption{Interference experiment: Laser light of frequency $\omega$ emitted by the source enters an interferometer through beam-splitter (half-silvered mirror) BS0 and gets detected after leaving beam-splitter BS1. The light can reach each of the detectors D(1) and D(0) by the paths $l$ and $s$; the path-length $l$ can be changed by the experimenter.}
\label{f1}
\end{figure}

Interference experiments are paramount in quantum mechanics. Consider the experiment sketched in Figure \ref{f1} using a Mach-Zehnder interferometer. For calculating the counting rates of each detector one must take into account information about the two paths leading from the laser source to the detector (\emph{wave behavior}). However, with a single-photon source only one of the two detectors clicks: either D($1$) or D($0$) (\emph{particle behavior}: ``one photon, one count'', or conservation of energy).

If $a \;\in\{+1,-1\}$ labels the detection values according to whether D($1$) or D($0$) clicks, the probability of getting $a$ is given by:
\begin{footnotesize}
\begin{eqnarray}
P(a)=\frac{1}{2}(1+ a \cos \mathit\Phi)
\label{Pa}
\end{eqnarray}
\end{footnotesize}
where $\mathit\Phi=\omega\tau$ is the phase parameter and $\tau=\frac{l-s}{c}$ the optical path.

By changing the phase parameter $\mathit\Phi$ (for instance by enlarging the length of path $l$) on gets different statistical distributions of the outcomes. Equal phases (mod. $2\pi$) define the same quantum state and yield the same distribution; different phases define different quantum states and yield different distributions, and different distributions correspond to different phases and quantum states.

To explain interferences the standard interpretation invokes the ``collapse of the wave function''. This rather cryptic expression means nothing other than the decision of the outcomes happens at detection. The ``collapse'' disposes of trajectories and is tacitly nonlocal \cite{as12}.
\vspace{0.2cm}

\noindent \textbf{``Collapse'' vs determinism}.\textemdash Determinism is the view assuming \emph{Principle A} and rejecting \emph{Principle Q}.

Determinism necessarily excludes any choice, specially free choice on the part of the experimenter: Otherwise (in the experiment of Figure \ref{f1}) the experimenter could access information about which of the two paths ($l$ or $s$) the particles takes after leaving BS0, and thwart the interference changing the other path. And even if one assumes that accessing which path information demolishes the particle, the experimenter could thwart interference according to the Equation (\ref{Pa}) by choosing  to change either path $l$ or path $s$ at will after the particle leaves BS0.

In the context of spin experiments, determinism is proved to conflict with quantum mechanics by the Kochen-Specker theorem \cite{ks}, unless one accepts that the particle's properties depend on the particular set of measurements the experimenter decides to perform (\emph{contextuality}).

In summary, determinism leads to \emph{superdeterminism}: In interference experiments (Figure \ref{f1}), the experimenter is contrived by nature to change the same path the particle takes at BS0, or alternatively the experimenter's path choice has back-ward effects determining the particle's path, and similarly for other properties.
\vspace{0.2cm}

\noindent \textbf{``Collapse'' vs locality}.\textemdash Assumed free choice, to explain interference one could in principle invoke a local version of standard quantum mechanics, by assuming decision at detection but excluding coordination between the detectors faster than light. In this sense this local version does not assume hidden variables. It has now been experimentally falsified by the experiment presented in \cite{gszgs, as12}, which thereby demonstrates the tacit nonlocality involved in the ``collapse'' in the context of single-particle experiments. Note that violation of Bell's inequalities cannot be implemented  in such experiments to prove nonlocality \cite{as12}.
\vspace{0.2cm}

\noindent \textbf{Alternative theories}.\textemdash These are explanations, assuming decision at the beam-splitter and rejecting both \emph{Principles A and Q}. Alternative theories assume \emph{accessible and inaccessible local hidden variables} and are sort of hybrids of the standard interpretation and the deterministic one. One can distinguish two types:
\vspace{0.2cm}

\emph{Type I}: Models yielding the same predictions as quantum mechanics, like Louis de Broglie's picture of the ``empty wave''. After leaving a beam-splitter, the \emph{accessible} particle travels always a well defined trajectory (say path $l$ in Figure \ref{f1}), and the \emph{inaccessible} ``empty wave'' (without energy and momentum) propagates through the alternative path (path $s$ in Figure \ref{f1}). In single-particle experiments the explanation escapes nonlocality at detection.

As far as one wishes to maintain locality, the explanation would conflict with quantum mechanics in 2-particle entanglement experiments, as proved by John Bell \cite{jb}. Nonetheless, the model can be extended to account for the quantum nonlocal correlations as well. To this aim the ``empty wave'' was supplemented by David Bohm with a ``nonlocal quantum potential'', and the so improved description yields the same predictions as the standard view in entanglement experiments with 2 or more particles (4 or more detectors) \cite{jb}.
\vspace{0.2cm}

\emph{Type II}: Models deviating from quantum mechanics. The three following have received extended attention:

- \emph{Eberhard}: assumes finite-speed causal influences propagating faster than light at a velocity $v$ ($c<v<\infty$) supposed to define a new constant of nature. The nonlocal coordination between the beam-splitters is supposed to break down when their separation is large enough (\cite{salart} and References therein).

- \emph{Suarez-Scarani}: assumes relativistic time-ordered nonlocal influences. The nonlocal correlations disappear when the beam-splitters are in a state of movement corresponding to a before-before relativistic configuration (\cite{zbinden, szgs} and References therein).

These two models include well defined experimental protocols leading to the disappearance of the nonlocal influences under certain conditions and thereby to local parts. The deviation from quantum mechanics is an axiom (to be tested), and not a theorem.

- \emph{Leggett}: combines as well nonlocal influences and local parts (\cite{as09} and References therein) but (contrarily to \emph{Suarez-Sacarani} and \emph{Eberhard}) \emph{Leggett} does not explain how the experimenter can thwart the nonlocal coordination in order to produce a set of local parts. The model pretends that deviation from quantum mechanics is not included in the main assumptions, but results as a theorem.

\emph{Suarez-Sacarani} and \emph{Eberhard} have been tested and falsified by experiment \cite{zbinden, szgs, salart}. More recently it has been proved that these two models lead to communication faster-than-light \cite{bancal, Scarani13}.

\emph{Leggett} has been ruled out by experiment (\cite{rc} and References therein). Notice however that this experiment does not refer to models assuming time-ordered nonlocal influences (like \emph{Bohm}, \emph{Suarez-Sacarani}, and \emph{Eberhard}) and therefore is not supposed to falsify these models. In this sense the experiment cannot be considered ``a complete answer to the question [...] of whether quantum mechanics is the optimal way to predict measurement outcomes.'' \cite{rc}

Obviously, \emph{nonlocal} alternative theories are no longer motivated by the wish of avoiding nonlocality. So the question arises why one still keeps to decision at the beam-splitter. The only plausible answer seems to be: to maintain trajectories and explain the outcomes in a deterministic way. But this is at odds with the incorporation of the free choice assumption, necessary to rule out locality. In this sense all the alternative theories have to be considered self-contradictory, unless they are susceptible of reformulation under the assumption of nonlocal decision at detection.
\vspace{0.2cm}

\noindent \textbf{``Many worlds'' and ``parallel lives''}.\textemdash The self-contradictory nature of the alternatives to quantum mechanics assuming hidden variables (i.e., rejecting the \emph{Principles A and Q}) is brought blatantly to light through the ``many worlds'' and ``parallel lives'' pictures.

Regarding ``many worlds'' Jon Bell said:

``The 'many world interpretation' seems to me an extravagant, and above all an extravagantly vague, hypothesis. I could also dismiss it as silly. And yet... It may have something distinctive to say in connection with the 'Einstein Podolsky Rosen puzzle', and it would be worthwhile, I think, to formulate some precise version of it to see if this is really so.'' (\cite{jb} p. 194).

Work by Lev Vaidman \cite{lev}, and more recently by Gilles Brassard and Paul Raymond-Robichaud \cite{brr}, shows that ``many worlds'' has really something distinctive to say in connection with the ``EPR puzzle'': Refutation of local hidden variables (by the violation of Bell's inequalities or other means) doesn't mean refutation of locality. By assuming decision at the beam splitter you become a ``strong Faithful'' of the ``Church of the Large Hilbert Space'' \cite{brr}, because (without even realizing it) you profess rejection of both \emph{Principle A} and \emph{Principle B}, and this rejection is the main article of the ``many worlds'' faith. Hence you will not have the necessary mental strength to reject locality.

It is not possible to keep decision at the beam splitter and nonlocality while rejecting ``many worlds'' in the name of freedom \cite{as12}:
\begin{small}
\begin{eqnarray}
&&\texttt{Decision at the beam-splitters}\;\& \; \texttt{Nonlocality}\nonumber\\
&&\Longrightarrow \texttt{Determinism}\nonumber
\end{eqnarray}
\end{small}
\indent What is more, any free choice comes from outside space-time to some extent. If decision of the beam-splitter is motivated by the desire of ``localizing'' the choice within space-time in order to deny \emph{Principle Q}, then the consequent attitude is to assume that there is no choice at all, that is ``many worlds''.

In summary, it is not by its own strength that ``many worlds'' lives, but by the weakness of its objectors.

By contrast if you assume nonlocal decision at detection you profess both \emph{Principle A} and \emph{Principle B}, and, therefore, you remain outside the Church of ``many worlds'' and will be able of opposing this interpretation without contradicting yourself. And you can also consistently reject superdeterminism for freedom's sake.
\vspace{0.2cm}

\textbf{Uncomplicated confutation of alternative theories}.\textemdash The previous analysis shows that strictly speaking Bell-type experiments demonstrate nonlocality only as far as one assumes decision at detection. One can certainly say that such experiments have demonstrated nonlocality between Alice's and Bob's detections. But it is also true that the detection loophole is not yet closed.

By contrast one can say that the experiment in \cite{gszgs, as12} demonstrates nonlocality in a more straightforward way, and (as discussed in \cite{as12}) without loopholes. Additionally, this experiment demonstrates that without nonlocality we could not have the most fundamental principle ruling the material world, the conservation of energy. Quantum physics has certainly to do with information, but fortunately also with physics.

Consider now what become the alternative theories under assumption of decision at detection:

The \emph{de Broglie-Bohm} theory  converts obviously into standard quantum mechanics.

As regards theories predicting disappearance of the nonlocal influences (\emph{Eberhard}, \emph{Suarez-Scarani}, and \emph{Leggett}), they lead to violation of the conservation of energy in each single quantum event and are falsified by the experiment presented in \cite{gszgs, as12} in a direct and uncomplicated way.

It is interesting to see that all these three models predict distributions depending on parameters other than the quantum mechanical phase values (like the parameter $\mathit\Phi$ in Figure \ref{f1}). In this sense they assume that a same quantum state can share two different distributions, and two different quantum states can share a same distribution, and can be considered representatives of the so (equivocally) called ``statistical interpretation'' \cite{pbr, rc12}. If one assumes decision at detection, the ``statistical interpretation'' amounts to assume nonlocality at detection together with at least two different distributions of detection outcomes depending on some (undefined) detector's characteristic (for instance in the experiment of Figure \ref{f1}). Therefore, by switching from one distribution to the other, one could signal from one detector to the other faster than light.
\vspace{0.2cm}

\textbf{``More nonlocal'' than quantum mechanics?}\textemdash What about possible alternatives that are more nonlocal than quantum mechanics and fulfill the no-signaling condition?

In \cite{as10a} it has been showed that the ``quantum algebra'' can be derived from three axioms with clear physical content: free will, conservation of energy, and locality emergent from nonlocality at detection. As far as one accepts these three axioms one has to accept Equation (\ref{Pa}) and exclude more nonlocality than Bell's one. Additionally if one keeps in mind that nonlocality at detection in single-particle experiments is the basic form of nonlocality, it is not clear what ``more nonlocal'' may mean in this context.
\vspace{0.2cm}

\textbf{Can quantum mechanics be improved?}\textemdash Does the previous conclusions mean that quantum mechanics is the ultimate theory and will not experience any further improvement in the future? Not by any means. Quantum physics has still to solve for instance the so called ``measurement problem'' (Schr\"{o}dinger cat paradox).

A reason for the little attention payed to nonlocality at detection so far, may be reluctance towards the ``subjective'' interpretation of the ``collapse'' as requiring the presence of a conscious observer. However, I would like to stress that it is possible to have a view combining the ``subjective'' aspect of Copenhagen and the ``objective'' one of GRW's ``spontaneous collapse'' \cite{gg} or Penrose's ``objective reduction'' \cite{rp}.

In fact, for measurement to happen it is not necessary at all that a human observer (conscious or not) is watching the apparatuses. However the very definition of measurement makes relation to human consciousness: An event is ``measured'', i.e. irreversibly registered, only if the objective conditions are fulfilled allowing a human observer to become aware of the registration.

In a sense I consider the ``collapse'' to be something as objective as ``death'', which physicians define as the \emph{irreversible} breakdown of all the brain functions including brainstem ones. For someone to die (generally) it is not necessary to be watched by some conscious physician. However the conditions defining ``death'' relate to the limit of the human capabilities to reverse a process of decay.

Even if measurement is basic to quantum mechanics, for the moment the theory does not define the conditions determining when the outcome gets \emph{irreversibly} registered and measurement happens (certainly, medicine does not achieve better in defining when precisely ``death'' happens). Ascertaining these conditions is to date an unsolved (but solvable) problem (``measurement problem''). This state of affairs  clearly shows a point where quantum theory, as we know it today, can and must be completed. And to do it, it may be that we have to understand better how consciousness and free will happen in the brain. I am convinced that the solution of this problem will bestow us a theory more fundamental than quantum mechanics, but not a more nonlocal one, or one renouncing to decision at detection.
\vspace{0.2cm}

\textbf{Conclusion}.\textemdash In 1997, together with Valerio Scarani, I proposed the before-before experiment, which became realized in 2001 \cite{szgs}. The people, my co-workers, were wonderful and the work was for sure enjoyable. However I wonder now why I proposed the before-before (and expended considerable work, time and money to do it) instead of proposing and doing the conceptually far more important and technically much less challenging experiment that I proposed in 2010 and has been done in the past months \cite{gszgs, as12}. A possible explanation may be that the new experiment is important not only because it is about nonlocality, but primarily because it demonstrates that nonlocality is crucial for the conservation of energy. To reach this insight, which now seems trivial to me, it was probably necessary to be defeated by quantum mechanics (in the field of the before-before) after having very much expected to beat it. Now I really understand how important decision at detection is.

The new experiment is also significant because it allow us to distinguish sharply quantum mechanics from alternative theories: If one ignores ``nonlocal decision at detection'', then the task of refuting alternative theories is rather complicate; if one assumes it, then the task seems rather trivial. Decision at the beam-splitter, or at detection, that is the question.

\end{document}